\documentclass[showpacs,12pt,preprintnumbers,amsmath,amssymb,floatfix]{revtex4-1}
\usepackage{amsmath}
\usepackage{graphicx}% Include figure files
\usepackage{dcolumn}% Align table columns on decimal point
\usepackage{bm}% bold math

\begin{document}
\title{A Free Energy Model of Boron Carbide}
\author{W. P. Huhn$^1$ and M. Widom$^1$}
\affiliation{$^1$Department of Physics, Carnegie Mellon University, Pittsburgh, PA 15213}

\date{\today}
\begin{abstract}
The assessed phase diagram of the boron-carbon system contains a single nonstoichiometric boron-carbide phase of rhombohedral symmetry with a broad, thermodynamically improbable, low temperature composition range.  We combine first principles total energy calculations with phenomenological thermodynamic modeling to propose a revised low temperature phase diagram that contains {\em two} boron-carbide phases of differing symmetries and compositions.  One structure has composition B$_4$C and consists of B$_{11}$C icosahedra and C-B-C chains, with the placement of carbon on the icosahedron breaking rhombohedral symmetry.  This phase is destabilized above 600K by the configurational entropy of alternate carbon substitutions.  The other structure, of ideal composition B$_{13}$C$_2$, has a broad composition range at high temperature, with rhombohedral symmetry throughout, as observed experimentally.
\end{abstract}
\maketitle

\section{Introduction}
\begin{figure}[b]
\includegraphics[width=0.75\textwidth, trim=0 0 0 2, clip]{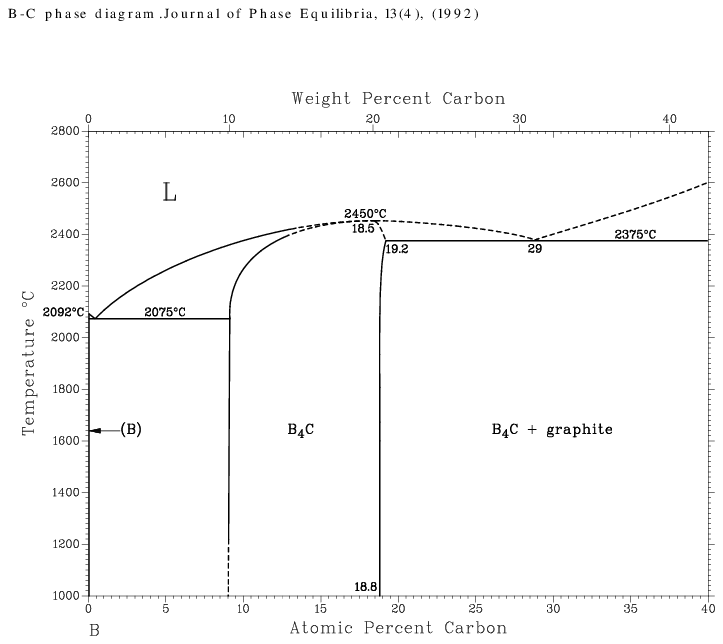}
\label{fig:exptPD}
\caption{Assessed phase digram of the boron carbide system~\cite{Okamoto92}.}
\end{figure}

The phase diagram of the boron-carbon system is highly controversial and has been frequently revised.  In addition to the pure elements, anywhere from one to eight different compound phases are claimed~\cite{Domnich11,SpringerMaterials}, with the sole point of agreement being the existence of a nonstoichiometric boron-rich rhombohedral phase known colloquially as ``boron carbide''.  Boron carbide is a hard material that is useful in making armor.  Owing to the neutron absorbance of boron, it also has uses as a shielding, control and shutdown material in nuclear power plants.  In the assessed phase diagram reflecting current consensus, reproduced in Fig.~\ref{fig:exptPD}, a single phase, labeled as B$_4$C and exhibiting rhombohedral symmetry, covers a composition range from a low carbon content of 9\% to a high of 19.2\%.  Note that the 20\% carbon content implied by the substance name B$_4$C is never achieved.

The broad composition range indicates a substitutional solid solution.  The temperature independence of the phase boundaries makes the assessed phase diagram thermodynamically improbable~\cite{Okamoto91}.  Specifically, if the phase field were to extend to absolute zero (T=0K), the positive entropy associated with substitutional disorder would imply a violation of the third law of thermodynamics~\cite{Abriata04}.  Several examples of such apparent violations are known, and generally are resolved by the onset of new phase behavior at low temperatures.  One famous example is the dispute over the phase diagram of plutonium-gallium~\cite{Hecker00}, where the American phase diagram extending the composition range of the $\delta$-phase to low temperature was eventually rejected in favor of the Russian version in which $\delta$ is stable only at elevated temperature.

Very likely, boron carbide is out of thermodynamic equilibrium at all but the very highest temperatures.  In this case, conventional experimentation cannot easily determine the equilibrium phase diagram, and theoretical approaches are needed to resolve the fundamental character of the stable state(s) at low temperature.  Previous first principles calculations utilizing electronic density functional theory~\cite{Mauri01,Saal07,Vast09,Widom12} indicate the existence of {\em two} boron-rich phases, one of ideal stoichiometry B$_{13}$C$_2$ and rhombohedral symmetry, the other of ideal stoichiometry B$_4$C=B$_{12}$C$_3$ and monoclinic symmetry.  In the present paper, we introduce a simplified thermodynamic model, inspired by first principles calculations, that elucidates the probable evolution of the equilibrium boron carbide phase field at low temperatures.

Crystallographic refinements~\cite{Clark46,Will76,Kwei96} claim that the B$_4$C phase has a 15-atom unit cell with rhombohedral symmetry.  These papers agree that the boron carbide structure can be viewed in terms of principal structural elements: boron icosahedra linked in a fashion similar to the $\alpha$-boron structure, and a linear chain of 3 atoms lying along the 3-fold axis of the unit cell. The atomic positions of the icosahedra belong to two site classes: polar sites linking the icosahedra to each other, and equatorial sites linking the icosahedron to the chains.

The distribution of carbon and boron atoms on these structural elements is uncertain.  An idealized B$_{13}$C$_2$ structure with rhombohedral symmetry occupies the twelve icosahedral sites with boron, while the three-atom chain takes the pattern C-B-C (i.e. the chain center atom is boron and the terminal sites are carbon).  However, first principles calculations~\cite{Bylander90,Mauri01} find that the most stable, enthalpy minimizing, structure is B$_{12}$C$_3$, with one carbon atom replacing boron on a polar site of the icosahedron, breaking the rhombohedral symmetry.  Experimentally~\cite{Schmechel2000}, the phase appears to contain a mixture of characteristic motifs: B$_{12}$ and B$_{11}$C icosahedra; C-B-C, B-B-C, B-V-B and C-V-C chains (V = vacancy).

The enthalpy minimizing B$_{12}$C$_{3}$ structure has monoclinic symmetry, in disagreement with the B$_4$C phase's observed rhombohedral symmetry.  We propose that the phase known as ``B$_{4}$C'' should be renamed ``B$_{13}$C$_{2}$'', or simply "rhombohedral", as we shall do for the remainder of this paper.  We also claim that there exists a second phase, which is the true B$_{4}$C phase, that we term ``monoclinic''.  In the monoclinic phase, all of the carbons lie on equivalent polar sites of the icosahedra, making a well-ordered structure.  Although there is an entropic term corresponding to the choice of this site, it is non-extensive and thus vanishes in the thermodynamic limit.  In the absence of substitutional disorder the monoclinic phase is a line compound.  In previous work~\cite{Widom12}, we proposed a phase transition in which the monoclinic phase transitions into the rhombohedral phase through the unlocking of a rotational degree of freedom in the placement of the polar carbons.  Note that swapping of carbon sites is equivalent to a rotation of a B$_{11}$C icosahedron.  Landau theory~\cite{Landau} predicts no such transition occurs in the rhombohedral phase, as the symmetry already matches the high T limit.  Here, we construct an analytic model to interpret our previous computational results, from which we can derive an actual phase diagram.

\section{Free Energy Model}
Four solid phases compete for stability: elemental boron ($\beta$-rhombohedral), elemental carbon (graphite), and monoclinic and rhombohedral boron carbide.  The first three phases are modeled as line compounds throughout the entire temperature range, but rhombohedral boron carbide will be allowed a carbon composition ranging from B$_{14}$C$_1$ ($x=1/15=0.067$) to B$_{12}$C$_3$ ($x=3/15=0.200$).  The stable T=0K composition for the rhombohedral phase is B$_{13}$C$_{2}$ ($x=2/15=0.133$).  According to first principles calculations~\cite{Widom12} the most stable B$_{14}$C$_1$ structure consists of B$_{12}$ icosahedra with B-B-C chains.  At composition B$_{12}$C$_3$ the most stable structure is monoclinic (i.e., not rhombohedral).  Rhombohedral structures at this composition correspond to B$_{11}$C icosahedra with C-B-C chains, but the placement of the carbon atom on the polar sites is randomly oriented among different primitive cells, unlike the monoclinic structure where all polar carbons are uniquely aligned throughout space.  Thus our model for the rhombohedral phase allows as structural units B12 and B11C icosahedra, and C-B-C and B-B-C chains.  Let $y_C$ be the fraction of icosahedra containing a polar carbon, and $y_B$ the fraction of chains containing a terminal B.  Note that $y_B$ and $y_C$ are bounded between 0 and 1, and that the carbon fraction
\begin{equation}
\label{eq:x}
x=\frac{1}{15}(y_C-y_B+2).
\end{equation}

\begin{figure}
\includegraphics[width=0.75\textwidth,angle=-90]{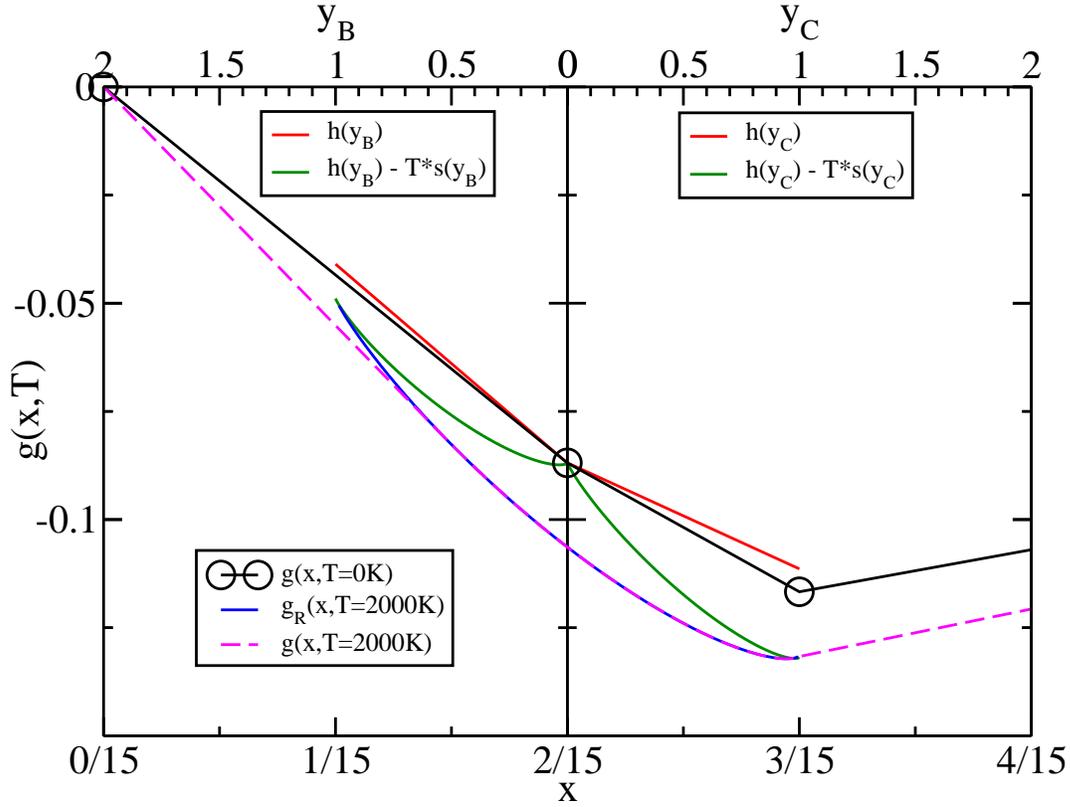}
\label{fig:gR}
\caption{Components of free energy model. Details in text.}
\end{figure}

The enthalpies of formation of $\beta$-rhombohedral boron and graphite vanish by definition, while we denote the T=0K enthalpies of the stable rhombohedral and monoclinic boron carbide phases, respectively, as $h_R^0$ and $h_M^0$.  Our first principles calculations yielded $h_R^0=-0.087$ and $h_M^0=-0.117$~ eV/atom (see Fig.~\ref{fig:gR}).  Notice that the monoclinic phase is more stable, at $x=0.200$, than the rhombohedral phase is at $x=0.133$.  We extend the rhombohedral phase entropy beyond its ideal composition by assigning an enthalpic penalty $\beta>0$ for each excess chain boron and an enthalpic benefit $\gamma<0$ for each icosahedral polar carbon.  Thus $h(y_B)=h_R^0+\beta y_B$ provided $y_C=0$ and similarly $h(y_C)=h_R^0+\gamma y_C$ provided $y_B=0$, as illustrated in Fig.~\ref{fig:gR}.  In general, $y_B$ and $y_C$ are both nonzero, so that
\begin{equation}
\label{eq:hR}
h_R(x)=h_R^0+\beta y_{B}+\gamma y_{C}.
\end{equation}

To complete the free energy model we assign an entropy based on random substitution disorder.  Specifically, an ideal site substitution entropy $-k_B(y\ln y + (1-y)\ln(1-y)$ for selecting the fraction $y$ of structural units on which to substitute, together with an intrinsic rotational entropy $k_B\ln 6$ for the orientation of the icosahedral polar carbon, and an intrinsic reflection entropy $k_B\ln 2$ for the choice of terminal chain position, as there are 6 polar sites per icosahedron and 2 terminal sites per chain. Thus,
\begin{multline}
\label{eq:sR}
s_R(x)=\frac{k_{B}}{15}(y_{B}\ln 2-y_{B}\ln y_{B}-(1-y_{B})\ln (1-y_{B})) \\
+\frac{k_{B}}{15}(y_{C}\ln 6-y_{C}\ln y_{C}-(1-y_{C})\ln (1-y_{C}))
\end{multline}
combines with the enthalpy eq.~(\ref{eq:hR}) yielding the free energy
\begin{equation}
\label{eq:gR}
g_R(x,T)=h_R(x)-T s_R(x).
\end{equation}
A similar free energy model was proposed by Emin~\cite{Emin88}, although he supplemented the free energy with an additional bipolaron density that is no longer considered as relevant.  As the monoclinic phase is a line compound, no composition dependence is required for its enthalpy, and because the rotational degree of freedom is locked, it lacks entropy.

Fig.~\ref{fig:gR} illustrates components of the free energy.  Black circles denote the enthalpies $h_R^0$ and $h_M^0$, while the black line segments are the convex hull of free energy $g(x)$ at $T=0K$.  Red lines and green curves illustrate enthalpy and $T=2000$K free energy under the constraint that either $y_B$ or $y_C$ vanishes, taking parameter values $\beta$=0.05 and $\gamma=-0.024$.  The blue curve illustrates the $T=2000$K free energy $g_R(x)$ (eq.~(\ref{eq:gR})) with the constraint relaxed, while the dashed pink line is the full convex hull of free energy $g(x)$.

The free energy $g_R(x)$ in eq.~(\ref{eq:gR}) is to be regarded as a Landau-type free energy, as the order parameters $y_B$ and $y_C$ must still be determined as a function of composition $x$ and temperature $T$.  Note the two order parameters are not independent, as the composition eq.~(\ref{eq:x}) implies $({\rm d}y_C/{\rm d}y_B)|_x=1$.  Now, minimizing $g_R$ with respect to $y_C$ yields
\begin{equation}
0=\left.\frac{dg_{rhom}}{dy_{C}}\right|_x=\beta+\gamma+\frac{k_{B}T}{15}\left(-ln12+ln\left(\frac{y_{B}}{1-y_{B}}\frac{y_{C}}{1-y_{C}}\right)\right)
\end{equation}
Substituting for the parameter $y_B$ and rearranging yields a quadratic equation for $y_C$,
\begin{equation}
\label{eq:yC}
y_{C}^2 (1-\kappa) + (2 - 15x(1-\kappa))y_{C}+\kappa(1-15x)=0
\end{equation}
where we define
\begin{equation}
\label{eq:kappa}
\kappa(T) \equiv 12 e^{-15(\beta+\gamma)/k_{B}T}
\end{equation}
as a measure of the extent to which polar boron atoms can swap positions with chain terminal carbons, which constitutes the fundamental excitation of the ideal rhombohedral structure.

\subsection{T $\rightarrow$ 0K limit}

We require that the enthalpy parameters $\beta+\gamma>0$ to ensure that the ideal rhombohedral structure with $y_B=y_C=0$ minimize the free energy at $x=2/15$ in the limit $T=0$.  Hence,
\begin{equation}
\label{eq:lim_kappa}
\lim_{T \to 0}\kappa(T)=0,
\end{equation}
reducing eq.~(\ref{eq:yC}) to the equation $y_{C}^2+(2-15x)y_{C}=0$ with two solutions: either $y_{C}=0$ (and then $y_B=2-15x$), or else $y_{C}=15x-2$ (and then $y_B=0$).  That is, either boron substitutes on chain terminal carbon sites, in which case $y_C=0$ and $x<2/15$, or else carbon substitutes on icosahedral polar boron sites in which case $y_B=0$ and $x>2/15$.  As a function of composition we have
\begin{equation}
\label{eq:xT0}
y_{B}=(2-15x)\theta(2/15-x),~~y_{C}=(15x-2)\theta(x-2/15)
\end{equation}
where $\theta$ is the Heaviside step function.

Although exact only at $T=0$K, the essential singularity in $\kappa$ makes this an excellent approximation over a wide range of temperature, up to 1000K and sometimes even higher in the examples discussed later.  The approximation represented by eq.~(\ref{eq:xT0}) separates the free energy into two branches, an $x < 2/15$ branch where only $y_{B}$ terms contribute, and an $x>2/15$ piece where only $y_{C}$ terms contribute, as illustrated in Fig.~\ref{fig:gR}. In each case, one branch of this piecewise-analytic free energy competes with a line compound, so we next derive a general equation for the phase boundary in this scenario.

\subsubsection{Substitutional Disorder Coexisting with a Line Compound}

To find the phase boundaries of rhombohedral boron carbide we must locate the coexistence of our substitutionally disordered phase with the competing phases.  Depending on composition and temperature the coexisting phase might be $\beta$-rhombohedral boron, monoclinic boron carbide, or graphite.  In every case the coexisting phase is treated as a line compound whose free energy is simply its enthalpy.  In this section we solve the coexistence equations generally, then apply this solution to specific phase boundaries in the following sections.

Consider the free energy model for substituting on a fraction $y$ of structural units, each with intrinsic multiplicity $\Omega$ and enthalpic penalty $\delta$,
\begin{equation}
\label{eq:g_model}
g(y,T)=h_R^0+\delta y - \frac{k_BT}{15}(y\ln\Omega-y\ln y-(1-y)\ln(1-y).
\end{equation}
Such a free energy represents one of the two branches of our rhombohedral free energy (eq.~(\ref{eq:gR})) at low temperature.  Let this phase coexist with an ordered line compound of free energy $g^*=h^*$ and ``composition'' $y^*$.  We now wish to find the composition, $y'$, of the disordered phase that coexists with the ordered phase at $y^*$.

Coexistence is determined by a double tangent condition, at $y'$ and $y^*$.  Specifically, there exists a straight line $f(y)=f_0+f_1y$ that is tangent to $g^*$ at $y=y^*$ and to $g(y,T)$ at $y=y'$.  For the line compound, tangency at $y^*$ is the simple condition $f(y^*)=h^*$.  For the disordered phase, tangency requires both that $f(y')=g(y',T)$ and that $f^\prime=f_1=g^\prime(y',T)$, where
\begin{equation}
g^\prime(y,T) = \delta-\frac{k_BT}{15}\left(\ln\Omega-\ln\frac{y}{1-y}\right).
\end{equation}
Solving, we find
\begin{equation}
\label{eq:yT}
\Omega^{y^*}e^{15(h^*-h_R^0-\delta y^*)/k_BT}=(y')^{y^*}(1-y')^{1-y^*}.
\end{equation}
In the case where $y' \approx 0$, this equation simplifies to
\begin{equation}
\label{eq:yT0}
y' \approx \Omega e^{15(h^*-h_R^0-\delta y^*)/y^*k_BT},
\end{equation}
while for $y'\approx 1$, we have
\begin{equation}
\label{eq:yT0p}
y' \approx 1-\left(\Omega e^{15(h^*-h_R^0-\delta y^*)/k_BT}\right)^{\frac{1}{1-y^*}}.
\end{equation}

\subsubsection{Boron-rich phase boundary}
\label{sec:Brich}
We apply this general solution to the specific case of rhombohedral boron carbide coexisting with $\beta$-rhombohedral boron.  In the notation of the preceding section, $h^*=0$.  There are two cases to consider depending on whether the phase boundary lies to the left or to the right of $x=2/15$.

{\bf Case 1.} If the boundary lies at $x\le 2/15$, so that $y_C=0$, then we identify $y=y_B$, $\delta=\beta$ and $\Omega=2$, as the disorder corresponds to substitution of boron onto the terminal chain carbon sites.  Also, $y^*=2$ corresponds to the composition $x=0$.  The requirement that $\beta$-rhombohedral boron be stable at $x=0$ against the boron carbide phase at $y_B=2$ implies a constraint that $2\beta>-h_R^0$.  The phase boundary occurs in the limit of small $y$, and from eq.~(\ref{eq:yT0}) we have
\begin{equation}
\label{eq:Bbound1}
y'_B=2 e^{-15(h_R^0+2\beta)/2k_BT}.
\end{equation}

{\bf Case 2.} If the boundary lies at $x\ge 2/15$, so that $y_B=0$, then we identify $y=y_C$, $\delta=\gamma$ and $\Omega=6$, as the disorder corresponds to substitution of carbon onto the icosahedral polar sites.  Also, $y^*=-2$ corresponds to the composition $x=0$.  The requirement that free energy be convex at $x=2/15$ implies a constraint that $2\gamma>h_R^0$.  The phase boundary occurs in the limit of small $y$, and from eq.~(\ref{eq:yT0}) we have
\begin{equation}
\label{eq:Bbound2}
y'_C=6 e^{+15(h_R^0-2\gamma)/2k_BT}.
\end{equation}
To determine if case 1 or 2 occurs, consider the ratio
\begin{equation}
y'_C/y'_B = 3 e^{-15(\gamma-\beta-h_R^0)/k_BT}.
\end{equation}
If $\gamma-\beta-h_R^0$ is positive, then the ratio vanishes at low temperature and the requirement of convexity places us in case 1.  If instead it is negative, then the ratio diverges and we have case 2.

\subsubsection{Carbon rich phase boundaries}

Next we apply our general solution to the specific case of rhombohedral boron carbide coexisting with monoclinic boron carbide.  Now, in the general notation, $h^*=h_M^0$.  As in the case of coexistence with elemental boron, two specific cases are possible.  However, the fact that carbon substitution is known to be energetically favorable (i.e. $\gamma<0$) implies that only the case of $y_B=0$ and $y=y_C$ is relevant.  Thus we identify $\delta=\gamma$ and $\Omega=6$.  Noting that the condition for low temperature stability of monoclinic B$_4$C against disordered rhombohedral at $x=3/15$ is $h_R^0+\gamma>h_M^0$, the phase boundary occurs in the limit of small $y'$, and from eq.~(\ref{eq:yT0}) we have
\begin{equation}
\label{eq:Cbound1}
y'_C=6 e^{-15(h_R^0-h_M^0+\gamma)/k_BT}
\end{equation}
as the phase boundary in coexistence with the monoclinic phase.

However, above a certain temperature $T_{0}$ the rhombohedral phase coexists with graphite.  We assume $T_{0}$ lies in the low temperature limit, and find the general form for the phase boundary due to coexistence between graphite and the rhombohedral phase.  In our general notation, $y^*=13$, $y=y_C$, $\delta=\gamma$ and $\Omega=6$.  In contrast to the preceding cases, the phase boundary occurs in the limit of $y\approx 1$, and from eq.~(\ref{eq:yT0p}) we have
\begin{equation}
\label{eq:Cbound2}
y'_C=1-6^{-\frac{13}{12}}exp(\frac{15(h_R^0+13\beta)}{12kT})
\end{equation}
as the phase boundary in coexistence with graphite.  The two carbon-rich boundaries cross at a certain temperature $T_0$ that can be determined by setting the values of $y'_C$ equal in eqs.~(\ref{eq:Cbound1}) and~(\ref{eq:Cbound2}).  For realistic parameters, the crossing occurs at $y_C\approx 1$.  Then from eq.~(\ref{eq:Cbound1}) we find
\begin{equation}
\label{eq:T0}
T_0\approx 15(h_R^0-h_M^0+\gamma)/k_B\ln 6.
\end{equation}

\subsection{Analytic T=$\infty$ Limit}
We now examine the high temperature limit for the free energy model.  For all possible values of $\beta$ and $\gamma$ we have $\lim_{T \to \infty}\kappa(T)=12$ giving the quadratic equation $-11y_{C}^2 + (2 + 165x)y_{C}+12(1-15x)=0$, only one of whose roots is physical: $y_{C}=\frac{1}{22}(2+165x-\sqrt{532-7260x+27225x^{2}})$.  Notice that only the multiplicities enter into this equation.  Although this equation is more complicated than the one obtained for the low temperature limit, it is defined across the composition range from $x=1/15$ to $x=3/15$.  The phase boundary on the boron rich side, $x_{boron}'$, is given by
\begin{equation}
\frac{dg_R}{dx}\mid_{x=x_{boron}'}=\frac{h_{\beta-boron}-g_R(x_{boron}',T)}{0-x_{boron}'}
\end{equation}
and the phase boundary on the carbon rich side $x_{carbon}'$ is given by
\begin{equation}
\frac{dg_R}{dx}\mid_{x=x_{carbon}'}=\frac{h_{graphite}-g_R(x_{carbon}',T)}{1-x_{carbon}'}.
\end{equation}
The high temperature limits are dominated by the entropic terms, yielding nonlinear implicit equations for $x'$.  Using a numeric equation solver we find $x_{boron}'=0.1095$ and $x_{carbon}'=0.1515$.  However these values are reached only at extreme high temperatures, while our model is intended only for use below the melting temperature.

\section{Realistic parameter values}

Our simple model depends on just four parameters, $h_R^0$, $h_M^0$, $\beta$ and $\gamma$.  Of these, the values of $h_{R,M}^0$ are easily determined from first principles calculations with simple idealized models.  Estimated values of $\beta$ and $\gamma$ may be obtained by inserting a single B or C substitutional defect into a hexagonal supercell of the ideal rhombohedral B$_{13}$C$_2$ structure.  A general description of the computational method is in Ref.~\cite{Widom12}.  Resulting values are listed in Table~\ref{tab:parameters}.  These parameters obey the constraints discussed in previous sections.  Because $\gamma-\beta-h_R^0=-0.0406$ is negative, we are in case 2 as discussed in Section~\ref{sec:Brich} where $x'_{Boron}>2/15$ at low temperature, although it eventually goes to $x'_{Boron}<2/15$ at very high temperatures.

\begin{center}
  \begin{table}
    \begin{tabular}{ | c | c | }
      \hline
      Parameter & Value (eV/atom)  \\
      \hline
      $h_R^0$ & -0.0869 \\
      \hline
      $h_M^0$ & -0.1167 \\
      \hline
      $\beta$ & 0.1031 \\
      \hline
      $\gamma$ & -0.0244\\
      \hline
    \end{tabular}
    \label{tab:parameters}
    \caption{Parameters for the rhombohedral phase obtained from first principles calculations.}
  \end{table}
\end{center}

There is some question whether our structural model is complete in the B-rich limit, as alternate structure models contain additional sites in the chain region, some only partially occupied~\cite{Yakel75,Morosin95,Kwei96}.  To take into account a possible influence of these additional sites, we investigate the effect of reducing the value of $\beta$, in order to model the effect of lower boron-rich enthalpies.  Note that increased multiplicity $\Omega$ would also enter the free energy linearly in $y_B$, though with an added factor of temperature $T$.

\begin{figure}
\includegraphics[width=0.8\textwidth,angle=-90]{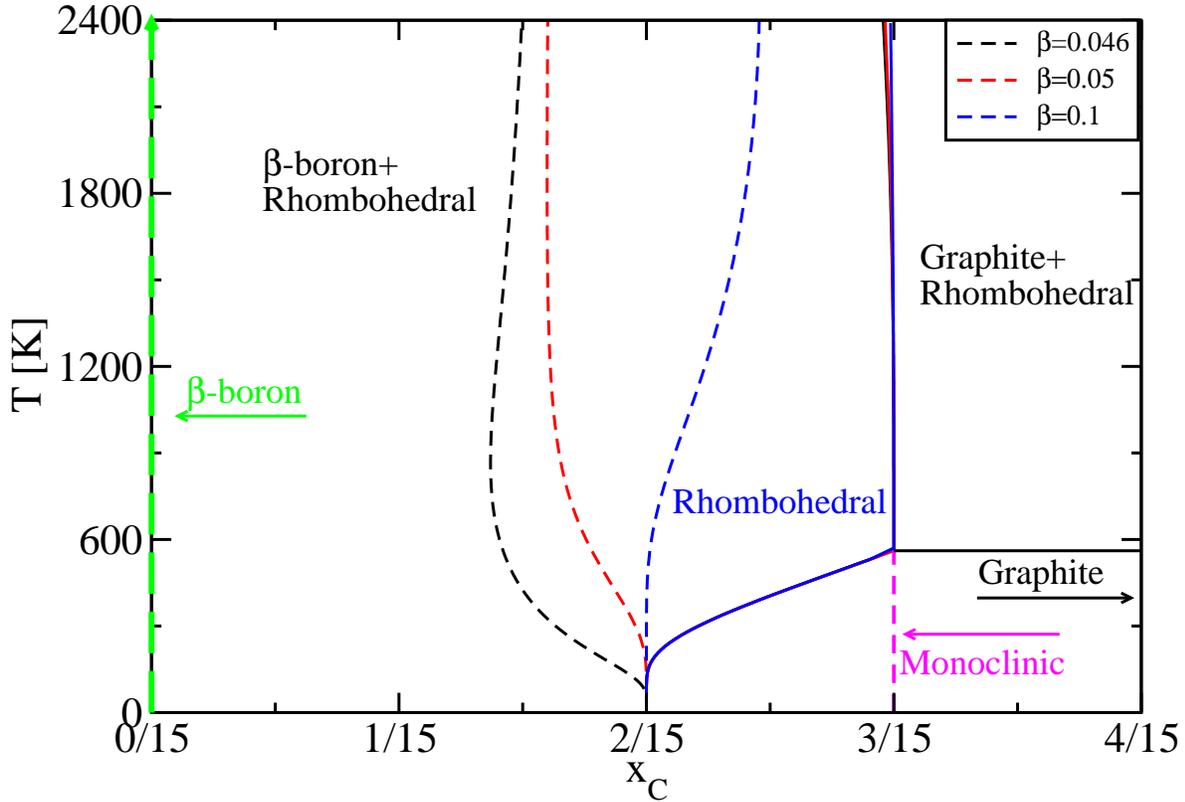}
\label{fig:calcPD}
\caption{Predicted phase boundaries for our model using calculated parameters as listed in Table~\ref{tab:parameters} and a selection of values of $\beta$.  Line compounds are $\beta$-boron at $x=0$, graphite (not shown) at $x=1$ and monoclinic boron carbide at $x=3/15$.  Carbon-rich phase boundaries of rhombohedral boron carbide are shown as solid lines, while boron-rich phase boundaries are shown as dashed lines.}
\end{figure}

Results for a selection of values of $\beta$ are shown in Fig.~\ref{fig:calcPD}. As expected, the monoclinic phase is destabilized above a temperature $T_0\approx600$K, at which it decomposes into a coexistence of carbon-rich rhombohedral phase together with graphite.  This value of $T_0$ is surprisingly consistent with the location of the heat capacity peak previously reported that was obtained from a completely different method~\cite{Widom12}.  The maximum carbon content of the rhombohedral phase is bounded below 20\%, owing to the logarithmic singularity in $s(y)$ at $y_c=1$ creating an infinite slope in $g(x,T)$ (to weak to be visible in Fig.~\ref{fig:gR}). Note that the Gibbs phase rule~\cite{Okamoto91} implies that rhombohedral and monoclinic boron carbide must have differing compositions while in coexistence with graphite, hence $x'_{C}=0.2000$ is forbidden in principle.

At temperature 2400K (around the melting point) the maximum carbon content depends on the value of $\beta$, and ranges from 19.9\% down to 19.7\% in our model for the given range of $\beta$ considered, whereas the experimentally assessed limit is 19.2\%.  The phase boundary in coexistence with boron depends strongly on the value of $\beta$, as can be seen in Fig.~\ref{fig:calcPD}.  The experimentally assessed limit is 9\% carbon, but over the range of $\beta$ values studied here, the limit ranges from 10-16\% carbon, strictly above the assessed value.  This is a further indication of inadequacy of our model in the boron-rich limit.

\section{Conclusion}
In this paper we propose a simple free energy model for boron carbide.  We predict the existence of two low temperature compounds, resolving the mystery of the assessed low temperature composition range.  One phase, whose ideal composition is $B_{13}C_{2}$, has rhombohedral symmetry throughout its wide high temperature composition range, consistent with experimental observations.  The other phase is a line compound of exact stoichiometry B$_4$C characterized by an array of parallel B$_{11}$C icosahedra whose symmetry is monoclinic.  Despite its favorable low enthalpy, this phase is predicted to be stable only below $T=600$K, a temperature so low that it might not be possible to form in thermodynamic equilibrium.  Various limiting behaviors of the model phase boundaries are derived analytically, including the key fact that $x'_{Carbon}<0.200$ in the rhombohedral phase.A quantitative discrepancy in the boron-rich limit reveals the need for further enhancement of our model.  Effects to consider include the additional interstitial sites, as well as electronic and vibrational~\cite{Shirai00} entropy.

\bibliography{bc}
\end{document}